\documentclass[%
 reprint,
 amsmath,amssymb,
 aps,
prb,
]{revtex4-1}
\usepackage{xcolor}
\usepackage{graphicx}
\usepackage{dcolumn}
\usepackage{bm}
\usepackage{mathrsfs}
\usepackage{hyperref}
\usepackage[normalem]{ulem}
\usepackage{lipsum}
\usepackage{IEEEtrantools}
\usepackage[version=3]{mhchem}
\usepackage{verbatim}
\usepackage{siunitx}

\begin{document}
\preprint{APS/123-QED}

\title{Determining Surface Phase Diagrams Including Anharmonic Effects}

\author{Yuanyuan Zhou}
\author{Matthias Scheffler}
\author{Luca M. Ghiringhelli}

\affiliation{Fritz-Haber-Institut der Max-Planck-Gesellschaft, Berlin-Dahlem, Germany}

\date{\today}

\begin{abstract}
We introduce a massively parallel replica-exchange grand-canonical sampling algorithm to simulate materials at realistic conditions, in particular surfaces and clusters in reactive atmospheres. Its purpose is to determine in an automated fashion equilibrium phase diagrams for a given potential-energy surface (PES) and for any observable sampled in the grand-canonical ensemble. The approach enables an unbiased sampling of the phase space and is embarrassingly parallel. It is demonstrated for a model of Lennard-Jones system describing a surface in contact with a gas phase. Furthermore, the algorithm is applied to Si$_M$ clusters ($M=\mathrm{2, 4}$) in contact with an H$_\mathrm{2}$ atmosphere, with all interactions described at the \textit{ab initio} level, i.e., via density-functional theory, with the PBE gradient-corrected exchange-correlation functional. We identify the most thermodynamically stable phases at finite $T, p(\ce{H_2}$) conditions.
\end{abstract}

\maketitle

\section{\label{sec:intro}INTRODUCTION}
  A prerequisite for analyzing and understanding the electronic properties and the function of surfaces is the detailed knowledge of the surface composition and atomistic geometry under realistic conditions. The structure of a surface at thermodynamic equilibrium with its environment is in fact a configurational statistical average over adsorption, desorption, and diffusion processes.
  \par A temperature-pressure phase diagram describes the composition and structure of a system at thermal equilibrium and is an essential tool for understanding material properties. The \textit{ab initio} atomistic thermodynamics (aiAT) approach \cite{weinert1986chalcogen, scheffler1, PhysRevB.35.9625, PhysRevB.38.7649, PhysRevB.58.4566} has been very successful in predicting phase diagrams for surfaces \cite{PhysRevB.68.045407, PhysRevLett.90.046103} and gas-phase clusters \cite{QUA:QUA24503, PhysRevLett.111.135501, 1367-2630-16-12-123016} at realistic \textit{T}, \textit{p} conditions. The key assumption is, however, that \textit{all} relevant local minima of the potential energy surface (PES) of a given system are enumerated, a (strong) limitation in case of unexpected surface stoichiometries or geometries. Such limitation can only be overcome by an unbiased sampling of configurational and compositional space. A further assumption in most work has been that the vibrational contributions to the change of the free energy are largely canceled and can be neglected. We will see below that this is not always justified.
  \par In this paper, we introduce a Replica-Exchange (RE) Grand-Canonical (GC) Monte-Carlo (MC)/Molecular-Dynamics (MD) algorithm, that enables the efficient calculation of complete temperature-pressure phase diagrams of surfaces, nanoparticles, or clusters in contact with reactive gas atmospheres. The RE and GC steps of the algorithm are formulated in the Metropolis MC framework, while the canonical sampling of configurations (diffusion) is supported via both MC and MD. In the case of surface in contact with a gas phase reservoir, the gas molecules can physi-/chemisorb on the surface, while adsorbed molecules or single atoms can desorb from the surface to the gas phase. At thermodynamic equilibrium , the number of desorbed molecules/atoms balances the adsorbed one, so that on average a constant number of molecules/atoms is present on the surface. We specifically target thermodynamically open systems in the GC ensemble, aiming at describing (nano)structured surfaces in a reactive atmosphere at realistic $T$, $p$ condition, so that the surfaces can exchange particles with the gas reservoir. The initial idea of RE \cite{PhysRevLett.57.2607, 0295-5075-19-6-002, doi:10.1080/01621459.1995.10476590, SUGITA1999141} is to allow for an efficent sampling of the configurational space by shuttling configurations from regions of low $T$ to regions of high $T$. Later, de Pablo \textit{et al.} \cite{doi:10.1063/1.480282, doi:10.1063/1.1456504} extended the concept to other intensive thermodynamic variables, such as the chemical potential ($\mu$)  in order to simulate the phase equilibria of Lennard-Jones (LJ) systems. This allows systems with different number of particles (the conjugate variable of $\mu$) to be shuttled across different values of $\mu$, thus enhancing the sampling, following the same spirit of the temperature replicas in traditional RE. By combining advantages of both GC and RE, our massively parallel algorithm requires no prior knowledge of the phase diagram and takes only the potential energy function together with the desired $\mu$ and $T$ ranges as inputs. The partition function is estimated using the output of the simulation, thus calculating thermodynamic observables is straightforward. 
    \par The structure of this paper is as follows. In Sec. \ref{sec:method} the method and  implementation  of our REGC  algorithm  will  be  discussed  in details. 
    In section \ref{sec:results} we show two applications of the REGC method. The first, in section\ref{ssec:lj}, proof-of-concept application is the determination of the $p$-$T$ phase diagram of a system composed of a LJ (frozen) surface in contact with a LJ gas phase. Next, in section \ref{ssec:si-clus}, we address the calculation of the phase diagram of the Si$_2$ dimer and Si$_4$ cluster in a reactive atmosphere of H$_2$ molecules by performing REGC with aiMD using Perdew-Burke-Ernzerhof (PBE)\cite{PhysRevLett.78.1396} xc approximation. 
    During the last several decades, silicon hydrides have attracted a lot of attention because of their potential applications in semiconductors, optoelectronics, and surface growth processes.\cite{sari2003mono, wang2001theoretical, xu1998photoelectron,kasdan1975laser} The binary clusters of silicon and hydrogen play key roles in the chemical vapor deposition of thin films, and photoluminescence of porous silicon. However, most of the previous research on silicon hydrides focused on the search of global minima structures, but the decisive issue of stability and metastability of silicon hydrides at realistic conditions (exchange of atoms with an environment) has not been addressed so far. The purpose of this application is to investigate the phase diagrams of silicon hydrides in reactive hydrogen atmosphere. 
    In the outlook section (\ref{sec:concl}), the capabilities and current limitations of our REGC method will be discussed.

\section{\label{sec:method}METHOD AND IMPLEMENTATION}
The sampling of complex systems, e.g., thermodynamically open systems, composed of many atoms arranged in molecules, clusters, condensed phases, etc., remains a challenge. 
The main factors that limit sampling efficiency is $(i)$ that systems' configurations get easily trapped --- especially at low temperatures --- in local minima and $(ii)$ the inherently long characteristic relaxation times in complex many-molecules systems (e.g., atoms' diffusion that require collective motions involving several degrees of freedom).
During the last decades, many powerful methods have been developed to deal with the first difficulty, e.g.,  J-walking \cite{doi:10.1063/1.458863, ORTIZ199866}, multicanonical sampling \cite{BERG1991249, PhysRevLett.68.9}, nested sampling \cite{PhysRevB.93.174108}, simple tempering \cite{doi:10.1063/1.462133, 0295-5075-19-6-002}, 1/k sampling \cite{PhysRevLett.74.2151}, expanded ensembles \cite{doi:10.1063/1.472257}, , and parallel tempering \cite{PhysRevLett.57.2607, SUGITA1999141}. While these methods are effective in overcoming kinetic barriers, they do little to ‘‘accelerate’’ the slow relaxation at low temperatures. \\
 Open ensembles, described at equilibrium by the grand-canonical-ensemble formalism, provide an effective mean to overcome slow-relaxation problems: atoms can get in and out of a system, effectively generating thermodynamically possible defects, along unphysical pathways (e.g., atoms' insertion or removal), thereby circumventing diffusional bottlenecks by disentangling degrees of freedom. We took advantage of both the replica-exchange and grand-canonical-ensemble concepts to design an algorithm that alleviate both kinetic trapping and slow phase space diffusion. In Sec. \ref{ssec:regc}, we describe our replica-exchange grand-canonical algorithm. Later, in Sec. \ref{sec:citeref} we describe how to use the results from replica-exchange grand-canonical simulations to calculate phase diagrams and free energy surfaces. 
 
\subsection{\label{ssec:regc}Replica-Exchange Grand-Canonical Monte Carlo / Molecular Dynamics}
\begin{figure}[ht]
    \centering
    \includegraphics[trim={0.02cm 0.0cm 0.02cm 0.0cm}, clip, width=0.56\textwidth]{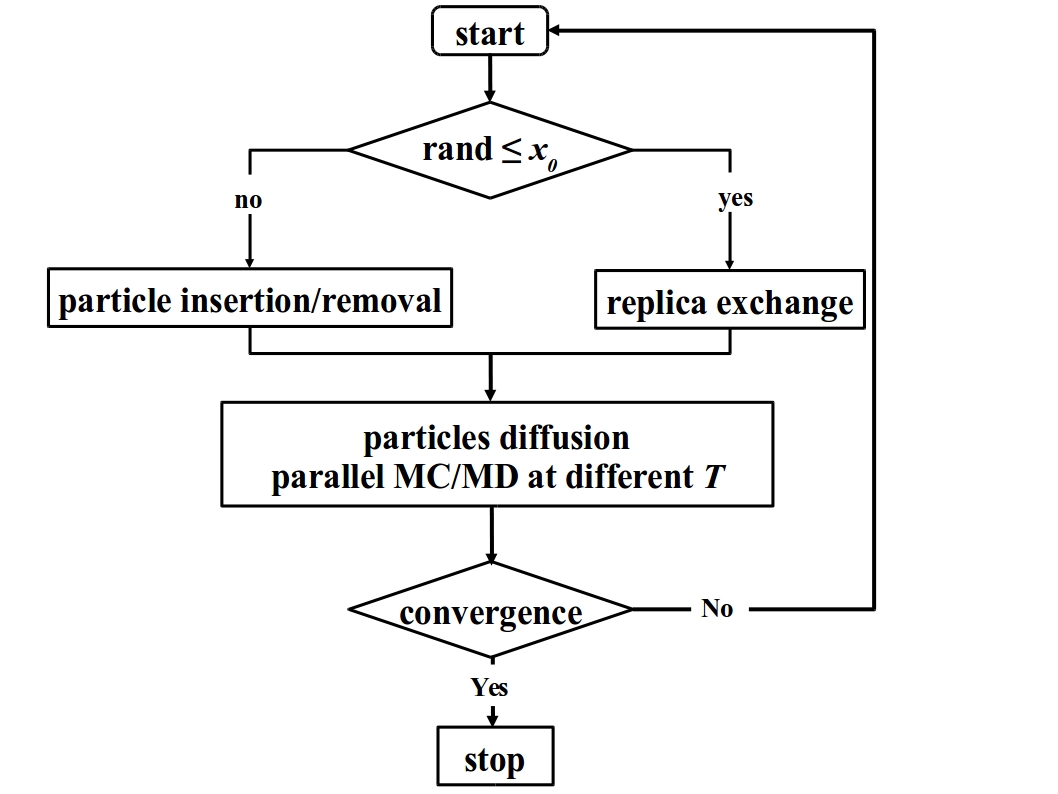}
    \caption{The flow chart of Replica-Exchange Grand-Canonical Monte Carlo/Molecular Dynamics algorithm. Here rand is a (pseudo) random number generated uniformly distributed between 0 and 1.}
    \label{fig:scheme}
\end{figure}

Our Replica-Exchange Grand-Canonical Monte Carlo or Molecular-Dynamics approach is outlined in Fig. \ref{fig:scheme}. In a REGCMC or REGCMD simulation, $S$ replicas of the original system of interest are considered, each evolving in a different thermodynamical states (${T}_{i}$ , ${\mu}_i$, where $i$ is the index of the replica). During the simulation, first the system has a probability $x_0$ $(0 \leq x_0 \leq 1)$ to attempt exchanging a particle with the reservoir and probability $(1-x_0)$ to perform a replica-exchange move (see below). After the particle/replica-exchange attempt, $S$ parallel molecular dynamics or Monte Carlo runs follow, to diffuse the system in the canonical ensemble. i.e., at temperature ${T}_{i}$, with fixed number of particles $N$ and volume $V$ of the system ($NVT$ ensemble). Then, the procedure is iterated until convergence of defined quantities is achieved. See further for the convergence criterion we adopted.

\subsubsection{\label{ssec:gc}Grand-Canonical Monte Carlo}
The particle insertion/removal step is handled by applying the formalism of the grand-canonical ensemble, where the subsystem of our interest (e.g., a surface or a cluster in contact with a gas phase), defined in a volume ($V$), is in equilibrium with a reservoir at given temperature ($T$), and chemical potential ($\mu$) of one  species (or more species, each with its own chemical potential). In practice, the reservoir is modeled as an ideal gas and $\mu$ depends on $T$ and the pressure $p$, as will be specified in the application cases. The number of atoms or molecules in the subsystem is a fluctuating variable, determined by specifying the chemical potential and temperature of the reservoir of (ideal) gas-phase atoms or molecules. The probability density of a grand-canonical ensemble of identical particles is:\cite{frenkel}

\begin{equation}
\label{Ngrand}
 \mathscr{N}_{\mu, V, T} (\bm{R};N) \varpropto \frac{e^{(\beta\mu N)}V^N}{\Lambda^{3N}N!} e^{[-\beta E(\bm{R};N)]}
\end{equation}
where $\beta= 1/k_\textrm{B}T$, $\Lambda = h/\sqrt{2 \pi m k_\textrm{B} T}$ is the thermal wavelength of a particle of mass $m$, and $E(\bm{R})$ is the potential energy of a configuration $\bm{R}$ of the $N$-particle system. The GCMC algorithm consists of the following MC moves: 1) insertion of a gas atom/molecule into the system at a random position, 2) removal of a randomly selected gas atom/molecule from the system, 3) displacement of a gas atom to a new random position in the system to sample the potential energy surface (PES). In our algorithm, the displacement (diffusion) is taken care of separately (see section \ref{sec:displ}) and can be done via either Monte Carlo or MD. Here, we consider the insertion and removal moves, where microscopic reversibility (also called `detailed balance', a \textit{sufficient} condition for an MC scheme to converge the evaluation of observable properties in the desired ensemble \cite{frenkel}) is ensured by having equal number of insertion and removal attempts, for all particles described by the given chemical potential. In practice, we first randomly select if a particle will be inserted or removed, i.e., by generating a (pseudo)random number $y^{\textsc{gc}}_1$ uniformly distributed between 0 and 1 and performing a removal if $y^{\textsc{gc}}_1<0.5$. 

For a removal, a particle (an atom or a molecule) is selected at random (by generating a new random number $y^{\textsc{gc}}_2$ and selecting particle $i$ if $(i-1)/N \leq y^{\textsc{gc}}_2 <i/N$). 
In order to fulfill detailed balance, a possible (and common) choice for accepting the removal of the selected particle is with probability\cite{frenkel}: 
\begin{equation}
 P_{(N \rightarrow N-1)} = \min\Big[1,\frac{\Lambda^{3}N}{V}e^{-\beta [\mu+E_{N-1}-E_{N}]}\Big]
\end{equation}
where $N$ is the number of atoms (or molecules) for which a reservoir at given temperature $T$ and chemical potential $\mu$ is defined, and which are in the system before the attempted removal. $E_{N}$ is the energy of the system of $N$ particles, $E_{N-1}$ is the energy of the same system, without the selected particle, and and $V$ is the system volume, which is fixed during the simulation. According to this formula, if the change in energy due to the particle removal is similar in value to $\mu$, there is a high probability that the removal is accepted.

For the insertion, first a location is randomly chosen, uniformly in the simulation volume (in a rectangular cell, by driving three independent uniformly distributed random numbers, one for each Cartesian coordinate). Then, a particle is positioned in the selected location and its insertion is accepted with probability \cite{frenkel}: 
\begin{equation}
 P_{(N \rightarrow N+1)} = \min\Big[1,\frac{V}{\Lambda^{3}(N+1)}e^{-\beta [\mu-E_{N+1}-E_{N}]}\Big]
\end{equation}
The probability of accepting an insertion can be low in dense systems as random locations will have high probability to end up too close to already-present particles, henceforth yielding large $E_{N+1}-E_{N}$ and consequent rejection of the insertion. Since we are modeling adsorption on surfaces or clusters in contact with a gas phase, we have a relatively rarefied system, especially if the considered volume of particle insertion (and removal) does not include the subsurface (see further).

\subsubsection{\label{ssec:re}Replica Exchange in the Grand-Canonical ensemble}
\begin{figure}[h]
  \centering
   \includegraphics[trim={0.06cm 0.06cm 0.02cm 0.05cm}, clip, width=0.6\textwidth]{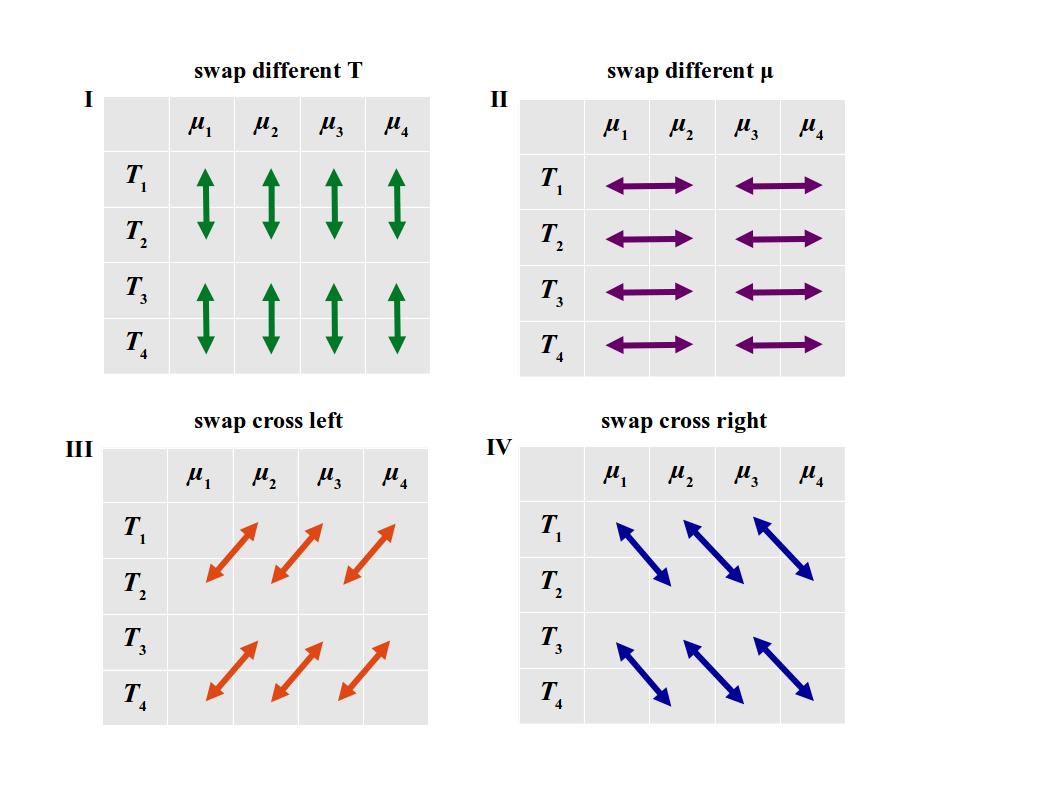}
  \caption{The 2D schematic of of  Replica-Exchange Grand-Canonical method.}
    \label{fig:mesh1}
\end{figure}

We define an extended ensemble that is the collection of $S = L \times M$ replica of a given system, arranged in $L$ values of temperature and $M$ values of the chemical potential, as illustrated in Fig. \ref{fig:mesh1}a. In this paper, we consider only one species that exchange particles with the reservoir, hence, one chemical potential. 
The \textit{partition function} of this extended ensemble is the product of the partition functions of the individual  $(\mu_{m},V,T_{l})$ ensembles, where $l = 1, 2, \ldots, L$ and $m = 1, 2, \ldots, M$: 

\begin{equation}
 Q_\textrm{extended} = \prod_{l=1}^{L}\prod_{m=1}^{M}\frac{e^{\beta_l \mu_m N_{l,m}} V^{N_{l,m}}} {\Lambda_l N_{l,m}!} \int d \bm{R} \, e^{-\beta_{l} E \left( \bm{R}; N_{l,m} \right) }
\end{equation}

In the following, we label the temperature indifferently by $T_l$ or $\beta_l = 1 / k_\textrm{B}T_l$.
The key observation is that taken one configuration along the evolution of a replica at given $(\mu_{m},V,T_{l})$, statistical mechanics allows us to write a well defined probability that the same configuration belongs to the another state $(\mu_{o},V,T_{k})$. 
We now randomly select a pair of replicas. The replica at state $(\mu_{m},V,T_{l})$ is in configuration $\bm{R}_i$ (e.g., represented by the $3\times N_{l,m}$ matrix of coordinates) and the replica at state $(\mu_{o},V,T_{k})$ is in configuration $\bm{R}_j$. We then aim at defining a rule for accepting the swap of the configurations between the two replicas, in order to satisfy the detailed balance in the extended ensemble. To the purpose, one has to impose the following equality:

  \begin{IEEEeqnarray}{Cl}
  \nonumber & \mathscr{N}_{(\beta_l, \mu_m, \bm{R}_i)}\mathscr{N}_{(\beta_k, \mu_o, \bm{R}_j)} \\ 
  \nonumber & \times P_{[(\beta_l, \mu_m, \bm{R}_i), (\beta_k, \mu_o, \bm{R}_j)\rightarrow (\beta_l, \mu_m, \bm{R}_j), (\beta_k, \mu_o, \bm{R}_i)]} \\ 
 \nonumber = & \mathscr{N}_{(\beta_l, \mu_m, \bm{R}_j)}\mathscr{N}_{(\beta_k, \mu_o, \bm{R}_i)} \\ 
  & \times P_{[(\beta_l, \mu_m, \bm{R}_j), (\beta_k, \mu_o, \bm{R}_i) \rightarrow (\beta_l, \mu_m, \bm{R}_i),(\beta_k, \mu_o, \bm{R}_j)]}
  \end{IEEEeqnarray}

where $\mathscr{N}$ is the probability density in the grand-canonical ensemble (Eq. \ref{Ngrand}), and $P$ is the probability to swap configurations.
Our choice of $P$ that satisfies the detailed balance is:
\begin{IEEEeqnarray}{Cl}
& \nonumber P_{[(\beta_l, \mu_m, \bm{R}_i)(\beta_k, \mu_o, \bm{R}_j)\rightarrow(\beta_l, \mu_m, \bm{R}_j)(\beta_k, \mu_o, \bm{R}_i)]} \\
= \nonumber &  \min \big [1, (\frac{\beta_l}{\beta_k})^{\frac{3}{2}(N_{l,m}-N_{k,o})} \times \\
 & \label{eq:Pacc} e^{[-(\beta_l-\beta_k)(E(\bm{R}_j)-E(\bm{R}_i)+(\beta_l\mu_m-\beta_k\mu_o)(N_{l.m}-N_{k,o})]} \big]
 \end{IEEEeqnarray}
A similar swap-acceptance probability has been proposed in Refs. \citenum{doi:10.1063/1.480282} and \citenum{doi:10.1063/1.1456504}, but we include a factor $(\frac{\beta_l}{\beta_k})^{\frac{3}{2}(N_{l,m}-N_{k,o})}$ that is probably neglected in those papers. Furthermore, our scheme adopts a two-dimensional grid of values of temperatures and chemical potentials, while in Refs. \citenum{doi:10.1063/1.480282} and \citenum{doi:10.1063/1.1456504} the values of $T$ and $\mu$ are constrained to be along a phase boundary of the studied system (vapor-fluid coexistence for the LJ system), therefore being a uni-modal scheme, i.e., one-dimensional in practice.

It is clear from Eq. \ref{eq:Pacc} that swap trial moves are more likely to be accepted the larger the overlap between the energy distributions of the two replicas. A large overlap of energy distribution is verified if the values of the thermodynamic variables $(\mu,T)$ defining the two replicas are not too dissimilar. In traditional one-dimensional RE, swap moves are attempted only between neighbor replicas. In that case, each replica has two neighbors (or one, for the largest an smallest values of the chosen replicated thermodynamic variable, typically $T$). In our two-dimensional scheme (Fig. \ref{fig:mesh1}), each replica has between 3 and 8 neighbors, thus enhancing the possibility for configurations to ``diffuse'' across replicas.
We adopted a ``collective'' scheme for the attempted swaps that involves the definition of four different types of neighboring swaps, as illustrated in Fig. \ref{fig:mesh1}. At each RE move, one type of swaps is selected at random (each with probability $1/4$). This choice has the advantage to involve all replicas (when the number of $T$-replicas and $\mu$-replicas is even) in one attempted swap. An alternative scheme could be to select randomly one replica and independently one neighbor to perform the attempted swap, then to repeat until no replica has an unselected neighbor. We are exploring this scheme for higher-dimensional settings (e.g., $T$ and more than one $\mu$ for more than one type of particles that are exchanged with the reservoir).

\subsubsection{Atoms' displacement} \label{sec:displ}
At each cycle of our REGC scheme, after the RE or GC move has been performed, the atoms in each replica perform in parallel a sampling of the canonical (fixed $N$, fixed $V$, fixed $T$) ensemble. This is achieved with the standard Metropolis MC or with MD. 

According to MC, one atom-displacement step requires to select at random one atom and assigning to it a random displacement, typically uniformly distribute in a cube or sphere of size comparable with the typical interatomic distances at equilibrium.
The move is accepted with probability\cite{frenkel}:
\begin{equation}
 P_{(\bm{r} \rightarrow \bm{r} + \Delta\bm{r})} = \min\Big[1,e^{-\beta [E(\bm{r} + \Delta\bm{r},\bm{r}^{N-1})-E(\bm{r},\bm{r}^{N-1})]}\Big]
\end{equation}
where $\bm{r}$ is the position before the random displacement $\Delta\bm{r}$ of the selected atom and $[E(\bm{r} + \Delta\bm{r},\bm{r}^{N-1})-E(\bm{r}^N,\bm{r}^{N-1})]$ is the potential-energy difference between the system with one atom displaced and all the other $N-1$ atoms kept in place, and the system before displacement.
In MC schemes, one cycle is the application of the attempted displacement $N$ times, so that on average each atom is attempted to be displacement once.

According to MD, the forces among atoms are calculated and the Newton equation is numerically integrated in order to obtain one displacement step for all atoms \cite{frenkel}. This scheme samples the constant energy, constant $V$, constant $N$ ensemble (microcanonical). In order to sample the canonical ensemble, the velocities of the atoms need to be modified in order to obey the Maxwell-Boltzmann distribution at the desired $T$. This is achieved via numerical thermostats \cite{frenkel}.

The choice between the two schemes, MC or MD, for the canonical sampling step of our REGCMC or REGCMD algorithm is dictated only by convenience. In both case our choice is to perform few (about 10) MD steps or MC cycles between two applications of the REMC step, in order to take full advantage of the enhanced sampling allowed by the REGC accepted moves.

\subsubsection{\label{sec:code}Implementation}
Due to the inherently parallel nature of replica exchange, the REGC method is particularly suitable to implement on super computers in parallel. MD or MC simulation of each replica at different $T$ are performed simultaneously and independently for the same time steps/MC moves. The whole computation resources are propotional to the number of replicas $S$, e.g., if each replica requires $q$ cores, in total, $S\times q$ cores are assigned to this REGC simulation.

\subsection{\label{sec:citeref}Calculating Phase Diagrams}

After a REGC simulation, we obtain $\Omega_{l,m}$ equilibrium samples from each of the $S=L \times M$ thermodynamic states $(\mu_m, V, T_l)$ within the grand-canonical ensemble. For each sample, a wide range of observable values can be collected, starting from the potential energy, the number of particles, and going to properties that are not related to the sampling rules. For instance, structural quantities like the radial distribution function or electronic properties such as the HOMO-LUMO gap of the system. 
In order to construct a phase diagram for the studied system, one has first to define which \textit{phases} are of interest. For instance, we can define as one phase all samples with the same number of particles $N$. The task is then to evaluate the free energy $f_i(\mu,T)$ of phase $i$, as function of $\mu$ and $T$, and for each value of $(\mu,T)$ the most stable phase is the one with lowest free energy.
From textbook statistical mechanics, the free energy is related to the probability $p_i$ to find the sampled system in a certain phase (i.e., having a certain value of an observable quantity) as follows:
\begin{eqnarray}
 \nonumber f_i(\mu,T) & = & -k_{\ce{B}}T \ln p_i(\mu,T) \\ \label{eq:textbook} &=& -k_{\ce{B}}T \ln \frac{\int_\Gamma d\bm{R} \, \chi_i(\bm{R}) \, q(\bm{R};\mu,\beta)}{\int_\Gamma d\bm{R} \, q(\bm{R};\mu,\beta)}
\end{eqnarray}
where $\bm{R}$ denotes the configuration of the system, $\chi_i$ is the indicator function for the state $i$ --- e.g., equal to 1 when $N$ is a given $N^*$ and 0 otherwise ---, and $q(\bm{R};\mu,\beta)$ is the density function for the specific statistical ensemble. The integrals are over the whole configuration space $\Gamma$. 

The normalization term at the denominator of Eq. \ref{eq:textbook} is known as the \textit{partition function}, $c(\mu,\beta)$. Once $q(\bm{R};\mu,\beta)$ is defined for the sampled ensemble (see further), the nontrivial task is to estimate $c(\mu,\beta)$, in order to evaluate the free energy and find its minimum. 

To efficiently estimate the partition function from our REGC sampling, we adopted the multistate Bennett acceptance ratio (MBAR)\cite{doi:10.1063/1.2978177} approach, as implemented in the pymbar code (\url{https://github.com/choderalab/pymbar}). The MBAR method starts from defining the reduced potential function for the grand-canonical ensemble $U(\bm{R};\mu,\beta)$ for state $(\mu,\beta)$~\cite{doi:10.1063/1.2978177}:
\begin{equation}\label{eq:grandpot}
U(\bm{R};\mu,\beta) = \beta \Big[E(\bm{R})-\mu N(\bm{R})\Big]
\end{equation}
where $N(\bm{R})$ is the number of particles for the considered configuration. We note that there is a sign mistake in front of $ \mu N $ for the corresponding formula in the original MBAR paper~\cite{ doi:10.1063/1.2978177privateComm}.
The grand-canonical density function is then $q(\bm{R};\mu,\beta) = \exp [ - U(\bm{R};\mu,\beta)]$.\\
The MBAR approach provides the lowest-variance estimator for $c(\mu,\beta)$, first by determining its value over the set of actually sampled states, via the set of coupled nonlinear equations \cite{doi:10.1063/1.2978177}:
\begin{equation}\label{eq:normlizationconstant}
 \hat{c}_{l,m} = \sum_{l=1}^L \sum_{m=1}^M \sum_{i=1}^{\Omega_{l,m}}\frac{q(\bm{R}_{i,l,m};\mu_m,\beta_l)}{\sum_{l=1}^L \sum_{m=1}^M \Omega_{l,m} \hat{c}_{l,m}^{-1}q(\bm{R}_{i,l,m};\mu_m,\beta_l)}
\end{equation}
where the index $i$ runs over all the samples in one state. Crucially, all samples enter the estimator for $\hat{c}_{l,m}$, at state $(l,m)$, irrespective of the state they were sampled in. Once the set of equations for the $L\times M$ $\hat{c}_{l,m}$'s is solved, $c(\mu,\beta)$ can be estimated for any new state $(\mu,\beta)$ via the same formula, with the observation that the $\hat{c}_{l,m}$'s at the denominator are now known.

Next, Eq. \ref{eq:textbook} can be evaluated. Following the example where the phase $i$ is identified by the number of particles in the system, the values of $N$ that minimizes $f_i(\mu,\beta)$ is the stable phase at the particular value of $(\mu,\beta)$. 
Graphically, one can assign a color to each value of $N$ and, for each $(\mu_i,\beta_j)$ on a grid, the color is assigned to a pixel of size $(\delta\mu,\delta\beta)$ centered at $(\mu_i,\beta_j)$ (see Fig. \ref{fig:pd1}).

In order to obtain a more familiar $(p,T)$ phase diagram from the evaluated $(\mu,\beta)$, we use the relationship $\mu(p,T)= k_{\ce{B}}T \ln (p/p_0)$, where $p_0$ is chosen such that $-k_{\ce{B}}T \ln (p_0)$ summarizes all the pressure-independent components of $\mu$, i.e., translational, rotational, etc. degrees of freedom. \cite{PhysRevB.68.045407,beret2014reaction,bhattacharya2014efficient}

We now turn our attention to evaluating the ensemble-averaged value of some property, at a given state point $(\mu,\beta)$. To give a concrete example for which we actually give results in section \ref{ssec:ljs}, let's consider the radial distribution function $g(r)$, i.e., the probability to find a particle at a given distance $r$ from any selected particles, averaged over all particles and samples. Here, we are in particular interested in the average (or expected) value of a property like $g(r)$ when the system is in a given phase, e.g., has a certain number of particles $N$. 
The ensemble average value of $g(r)$ at a given $r$ and given state point $\mu,\beta)$, and phase $i$ is:
\begin{equation}
\langle g(r) \rangle _{\mu,\beta,i}  =   \frac{\int_\Gamma d\bm{R} \, \chi_i(\bm{R}) \, g(r;\bm{R}) \, q(\bm{R};\mu,\beta)}{\int_\Gamma d\bm{R} \, q(\bm{R};\mu,\beta)}
\end{equation}
where the function $g(r;\bm{R})$ at any given $r$ depends on the whole configuration $\bm{R}$.
In the MBAR formalism, the integrals are estimated over the sampled points via:
\begin{equation}\label{eq:avegr}
\langle g(r) \rangle_{\mu,\beta,i}  =  \sum_{n=1}^{\Omega_i} \frac{g(r;\bm{R}_n) \, c_{\mu,\beta}^{-1} \, q(\bm{R}_n;\mu,\beta)}{\sum_{l,m} \Omega_{l,m,i} c_{\mu_m,\beta_l}^{-1}q(\bm{R}_{l,m,i};\mu_m,\beta_l)} 
\end{equation}
where $\Omega_i$ is the number of samples in phase $i$ and therefore the sum over $n$ runs over all samples belonging to phase $i$. Similarly, $\Omega_{l,m,i}$ is the number of samples in phase $i$ in each sampled state point $(m,l)$.
In practice, $g(r)$ is discretized into a histogram, in which bin $k$ counts how many particles are found between distance $r_{k−1}$ and $r_k$ (see section section \ref{ssec:ljs} for more details). One should note that the average value of each bin in the histogram is evaluated independently by MBAR.

\section{\label{sec:results}RESULTS}
\subsection{\label{ssec:lj} Lennard-Jones surface}
As first example, we applied our REGC algorithm to a two-species Lennard-Jones (LJ) system, consisting of a fcc(111) frozen surface of species A, in contact with a gas phase of species-B particles. Details on the  interactions between BB and AB LJ particles are given in the Appendix, here we mention that we chose them so that AB interactions are much stronger than BB (being the A particles forzen, there is no interaction defined among them). The equilibrium distances $d_{ij}^\textrm{eql}$ are mismatched such that $d_{AB}^\textrm{eql} > d_{BB}^\textrm{eql}$, and both are shorter than the fixed AA first-neighbor distances. Other choices are possible, but here we focus on only one choice, in order to show in depth the type of \textit{a posteriori} analysis an REGC run allows for. The sub-system labeled as A$_{18}$ is a 2-layer slab with a $ 3 \times 3$ lateral supercell (i.e., 18 A atoms), periodically replicated in the $x$ and $y$ direction, while the $z$ direction is aligned with the [111] direction of the slab. The gas particle B is only allowed to insert in the ``surface" zone. We defined the ``surface" zone as a slab of height 48.0 {\AA} above (i.e., in the positive $z$ direction), starting from the $z$ position of the topmost atoms of A$_{18}$. At the same time, particles B are inserted at all $x$ and $y$ coordinates, uniformly. Insertion and deletion attempts have been performed with equal probabilities. Ten sequential MC moves are performed after each particle/replica exchange attempt. 
\begin{figure}[h!]
    \centering
    \includegraphics[trim={4.5cm 2.5cm 6.5cm 0.5cm}, clip, width=0.5\textwidth]{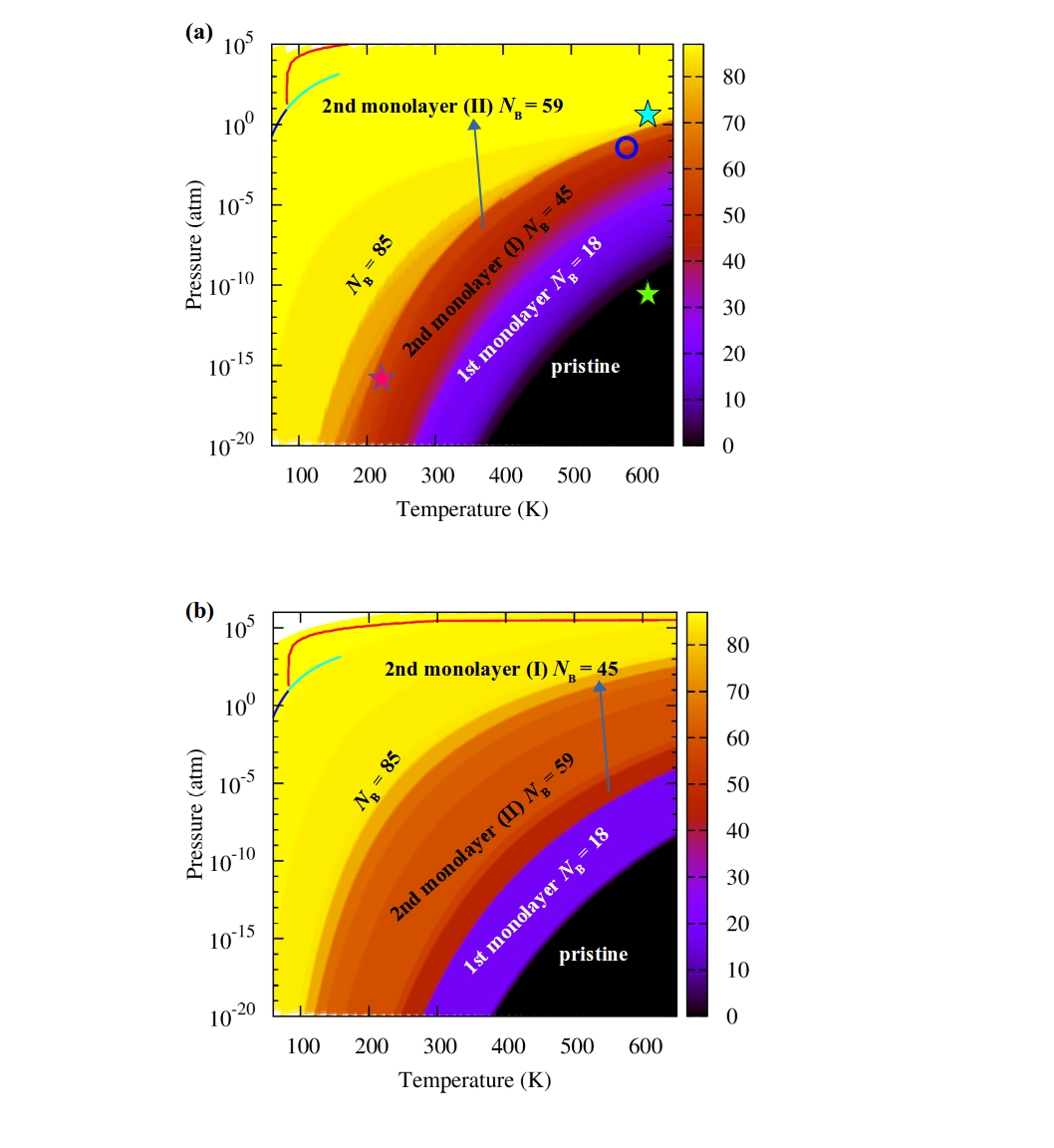}
\caption{Phase diagrams of a LJ gas-phase (particles B) in contact with a frozen fcc(111) LJ frozen surface calculated via by MBAR from the REGCMC sampling (panel a) and aiAT (panel b) at $(p_\textrm{B},T)$ conditions corresponding to a range from zero adsorbed particles (all in gas phase, region labeled as ``pristine'', referred to the surface) to the deposition of the LJ B particles into a bulk solid. The red line is the melting line for the LJ B particles, the sublimation line is blue, and the vaporization line is cyan. The cyan, green and pink stars correspond to the ``corner'' states for the REGCMC sampling: (650 K, -0.9 eV), (650 K, -2.4 eV) and (200 K, -0.9 eV), respectively. The fourth corner, (200 K, -2.4 eV) falls outside the $(p,T)$ window shown in the plot. The blue circle indicates (600 K, \num{8.89e-2} atm) and (200 K, \num{2.03e-17}atm) is exactly the pink star, corresponding to two states in Fig. \ref{fig:lj-rdf2}c and Fig. \ref{fig:lj-rdf2}b, respectively.}
    \label{fig:pd1}
\end{figure}
In the calculations, 160 replicas are defined i.e., 10 temperatures ranging from 200 to 650 K, with an interval of 50 K, and 16 chemical potentials ranging from -2.4 to -0.9 eV, with an interval of 0.1 eV. The range of chemical potentials is selected such that the lowest value of $\mu$ is comparable to and slightly lower than the adsorption energy of one B particle on A$_{18}$, in order to assure that the sampling includes states where zero or few particles are adsorbed (in order to have the pristine surface appearing in the phase diagram). The highest value of $\mu$ is ideally always close to zero, in order to scan up to the condensation of B particles and formation of a bulk B phase. The range of temperature was chosen to be slightly lower than the solid/liquid/gas triple point of the B particles and ranging to few times (here, four) its critical temperature \cite{doi:10.1080/00268977600100281}. In practice, pre-knowledge of the studied system can be applied in order to frame a suitable ($\mu,T$) window containing phases of interest. The spacing between $T$ and $\mu$ values is more difficult to estimate \textit{a priori}. During the simulation, one has to check that the acceptance ratio of RE attempted moves is not too low, in order to ensure a proper diffusion of replicas in the $(\mu,T)$ window. For instance, the present choice ensured an acceptance ratio of about 25\%.
Configuration swaps were attempted every 100 REGC steps, and $x_0$ was set equal to 0.99; a total of $1.2 \times 10^5$ REGC steps were performed to reach convergence, that is, there was no change in the density of reduced-energy states $\rho(U)$, with increasing simulation steps. The density $\rho(U)$ is sampled by binning the sampled configurations according to their value of $U$.

\subsubsection{\label{ssec:LJpd}Phase diagram}
The phase diagram shown in Fig. \ref{fig:pd1}(a) is constructed by using MBAR and shows the $(p_\textrm{B},T)$ regions where different number of adsorbed B particles are in thermodynamic equilibrium with their gas phase. The B reservoir is assumed to be an ideal gas, so the chemical potential of the reference state is defined as $\mu^0_{id.gas}\equiv k_{\ce{B}}T\ln(\Lambda^3)$. The relationship between pressure $p_{id.gas}$ in the reservoir and the chemical potential $\mu$ is $\beta\mu\equiv \beta\mu^0_{id.gas}+\ln(\beta p_{id.gas})$, that is $p_0=(k_{\ce{B}}T)^{\frac{5}{2}}(\frac{2\pi m}{h^2})^{\frac{3}{2}}$. The whole output data of REGCMC is sub-sampled every 100 REGC steps, that is, recording data after every attempted replica exchange, to remove correlations in the sampled quantities.

\begin{figure}[hb!]
    \centering
    \includegraphics[trim={9cm 4.5cm 10cm 2cm},  width=0.35\textwidth]{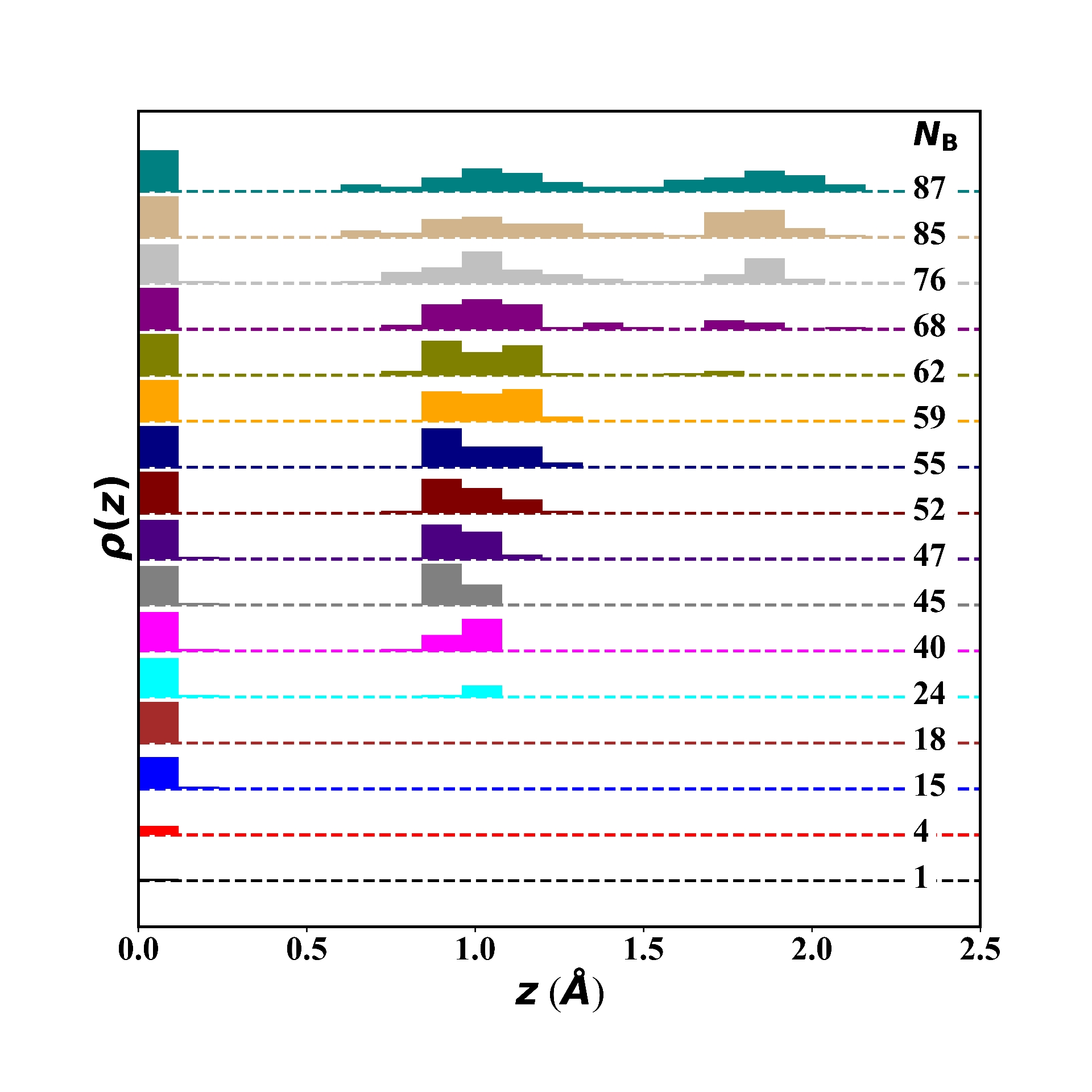}
    \caption{Axial distribution function of adsorbed particles for each $N_\textrm{B}$ composition generated in REGC sampling. The curves are displaced by 20 units and each dash line is a zero reference line for the curve with the same color.}
    \label{fig:lj-sdf}
\end{figure}

The MBAR@REGC phase diagram is compared to the aiAT@REGC phase diagram (Fig. \ref{fig:pd1}(b)), which is calculated via the following steps: $(i)$ For each observed number $N_\textrm{B}$ of adsorbed (B) particles in the REGCMC sampling, the lowest energy configuration is selected. We note that identifying phases (the phase is identified by $N_\textrm{B}$) via grand-canonical sampling is not the usual strategy for aiAT. Typically phases are enumerated on the basis of pre-knowledge and local minimization (at fixed number of adsorbed particles. In other words, the aiAT study presented in this paragraph is already richer than usual due to the unbiased structure sampling. $(ii)$ The formation Gibbs free energy for each of these phases is calculated via: 
\begin{equation}\label{eq:aiTherm}
\Delta G^f_{N_\textrm{B}}(T,p_B) = F_{N_\textrm{B}} - F_{\textrm{A}_{18}}-N_\textrm{B}\mu(T,p_\textrm{B})
\end{equation}
Here, the free energy of $F_{N_\textrm{B}}$ of the system $\textrm{A}_{18}\textrm{B}_{N_\textrm{B}}$ and $F_{\textrm{A}_{18}}$ of the pristine $\textrm{A}_{18}$ slab is approximated by the LJ energies of the two systems, i.e., all the vibrational contributions to the free energy are assumed to cancel out. This is often a justified assumption for systems studies via aiAT\cite{PhysRevB.68.045407}. As we will see, it is not a good approximation for this LJ system, at least at larger $N_\textrm{B}$. $(iii)$ As for MBAR@REGC, at each $(T_i, \mu_j)$ on a grid the phase with the lowest $\Delta G^f$ determines the color of the pixel of size $(\delta T,\delta\mu)$ centered at $(T_i,\mu_j )$. This aiAT@REGC approach, used here only for comparing to MBAR@REGC in order single out the role of the vibrational contribution to free energy, including anharmonic effects, is similar to the method recently proposed in Ref.~\onlinecite{doi:10.1021/acs.jpcc.8b11093}. There, the configurations are sampled by means of an approximated GC scheme at one temperature only and without replica exchange for either temperature or chemical potential. The effect of the reservoir to the free energy is taken care of by an expression similar to Eq. \ref{eq:aiTherm}.\\ 
By comparing the two panels of Fig. \ref{fig:pd1}, we note that up to $N_\textrm{B}\,=\,18$, the two phase diagrams almost coincide, especially at lower temperatures (in the Suppl. Material, we show a zoom-in of the region between 60 and 350 K). There are, however, significant differences at larger $N_\textrm{B}$: There are many more phases in Fig. \ref{fig:pd1})(a) that are missing in Fig. \ref{fig:pd1})(b) for $N_\textrm{B}>18$ and the region of stability of larger coverages is shifted to higher temperatures and lower pressures. This can be understood as due to increasingly larger vibrational contributions, especially in the direction $z$, perpendicular to the slab, while at low coverage the free energy is indeed essentially given by the LJ energy. We come back to this in the next section, after analyzing the structural properties of the different phases.\\
The analysis of the phase diagram Fig. \ref{fig:pd1})(b) reveals that for many values of number of adsorbed B particles, $N_\textrm{B}$, there is a region of stability in the phase diagram, however, for some specific values of $N_\textrm{B}$ larger stability areas are found. Besides $N_\textrm{B}\,=\,0$(the pristine surface), we recognize $N_\textrm{B}\,=\,18$ as the first complete mono-layer, $N_\textrm{B}\,=\,45$ as the addition of a second complete monolayer, plus a third phase, $N_\textrm{B}\,=\,59$ with a thicker second monolayer (see further). We also identify a large-coverage phase, $N_\textrm{B}\,=\,85$ which can be described by the formation of a ``third'' layer around 1.9 {\AA}, but in this case the particle distribution does not go completely to zero between second and third layer as it does between first and second, as shown in Fig. \ref{fig:lj-sdf}.
The diagram extends till the melting (red), vaporization (cyan), and sublimation (blue) line for bulk B particles. The phase transition curves are derived from the published equations of state for the LJ system.\cite{nicolas1979equation, johnson1993lennard, kolafa1994lennard, tang1999phase, okrasinski2001mathematical, van2000free, van2002gas} We underline that the phase diagram outside the $(p,T)$ region sampled directly via the REGC run is not \textit{extrapolated}. It is obtained as for all the diagram by Boltzmann re-sampling the configurations actually visited, using the measured (reduced) potential energies. 

\subsubsection{\label{ssec:ljs}Structural properties}
\begin{figure}[h]
    \centering
    \includegraphics[trim={4.3cm 5.0cm 6cm 0.9cm}, clip, width=0.5\textwidth]{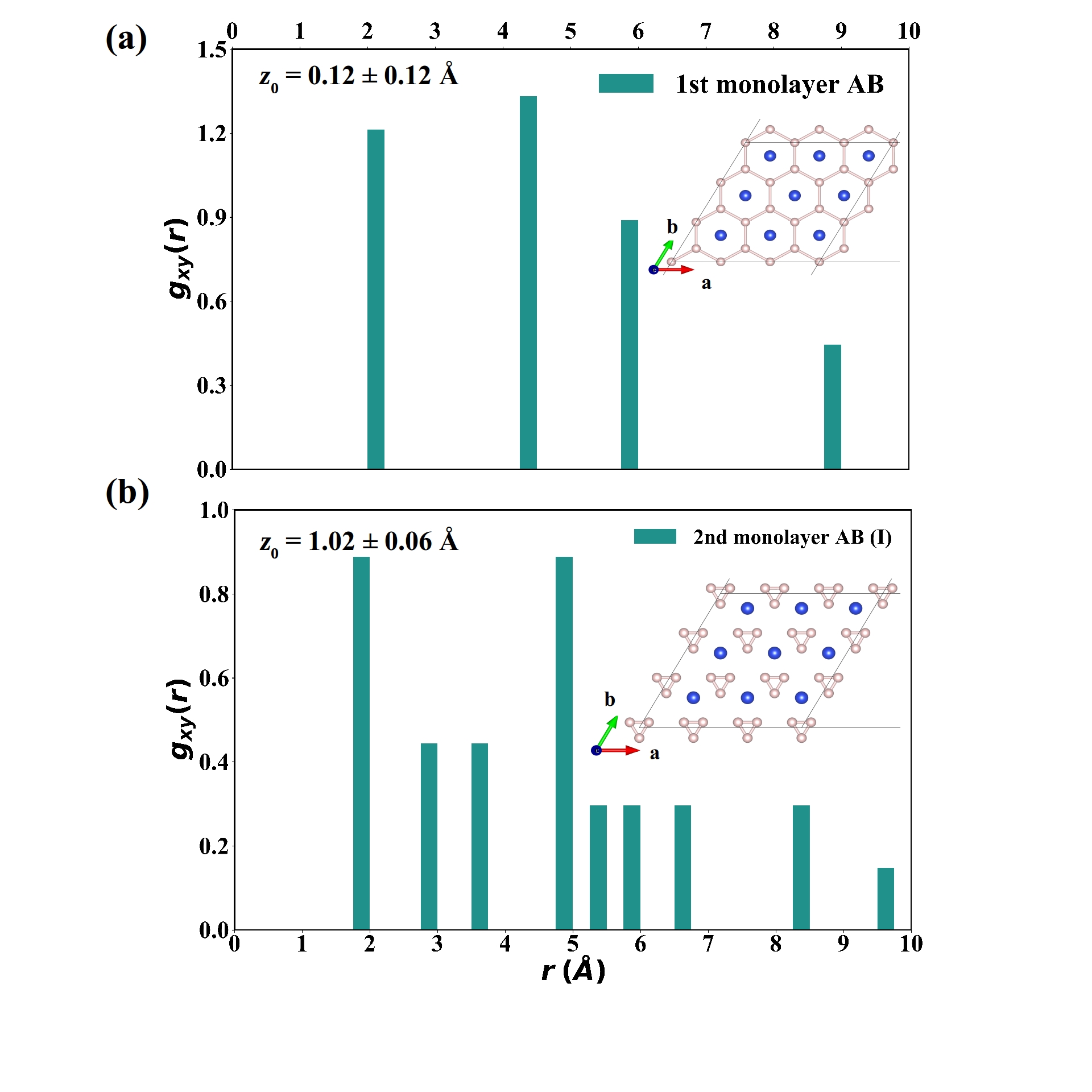}
    \caption{Lateral radial distribution functions $g_{xy}(r)$ for (a) first monolayer, and (b) second monolayer (I), respectively. The blue and pink balls in the insets indicate A and B particles, respectively.}
    \label{fig:lj-rdf1}
\end{figure}

The REGC sampling allows for much deeper analysis than the evaluation of the phase diagram. For instance, the structural properties of the adsorbed phases can be characterized in a statistical way.
The axial distribution function $\rho(z)$ was calculated by dividing the cell into slabs of width 0.12 {\AA}, parallel to the surface, and collecting a histogram of the number of particles in each slab along the REGC sampling. As shown in Fig. \ref{fig:lj-sdf}, the adsorbate has a clear layered structure up to the second layer. For larger $N_\textrm{B}$, i.e., $N_\textrm{B} > 59$, there are more and more particles adsorbed in the range $1.2\! \le z \! \le 1.8$ \AA, though another noticeable peak around 1.9 {\AA} occurs. As intuitively predictable, the first layer consists of 18 B particles located in all the hollow sites of the $3\times 3$ surface. When the second full monolayer $N_\textrm{B}\,=\,45$ is stable, the B particles occupy the 27 bridge sites of the A$_9$ surface layer. 
\begin{figure}[h]
    \centering
    \includegraphics[trim={8.cm 0cm 11cm 0.2cm}, clip, width=0.5\textwidth]{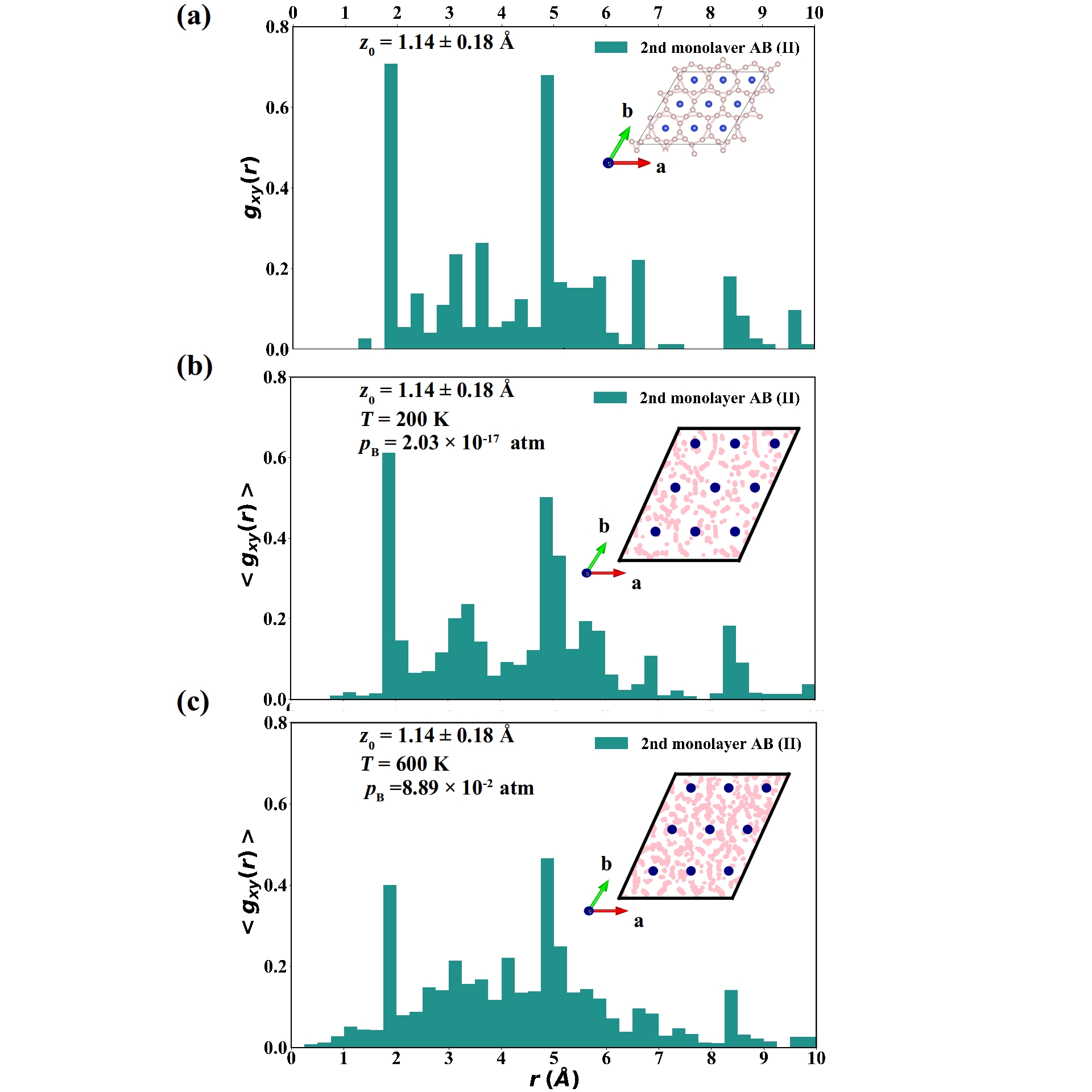}
    \caption{Lateral radial distribution function $g_{xy}(r)$ for (a) relaxed second monolayer (II) $\textrm{A}_9\textrm{B}_{59}$, average distribution function $<g_{xy}(r)>$ at (200 K, $2.03 \times 10^{-17}$ atm) state (b), and at (600 K, $8.89 \times 10^{-2}$ atm) (c) for the same composition. The blue and pink balls in the insets indicate A and B particles, respectively. }
    \label{fig:lj-rdf2}
\end{figure}

\begin{figure*}[ht]
    \centering
	\includegraphics[trim={0.8cm 3.0cm 0cm 2.2cm}, width=\textwidth, height=10cm]{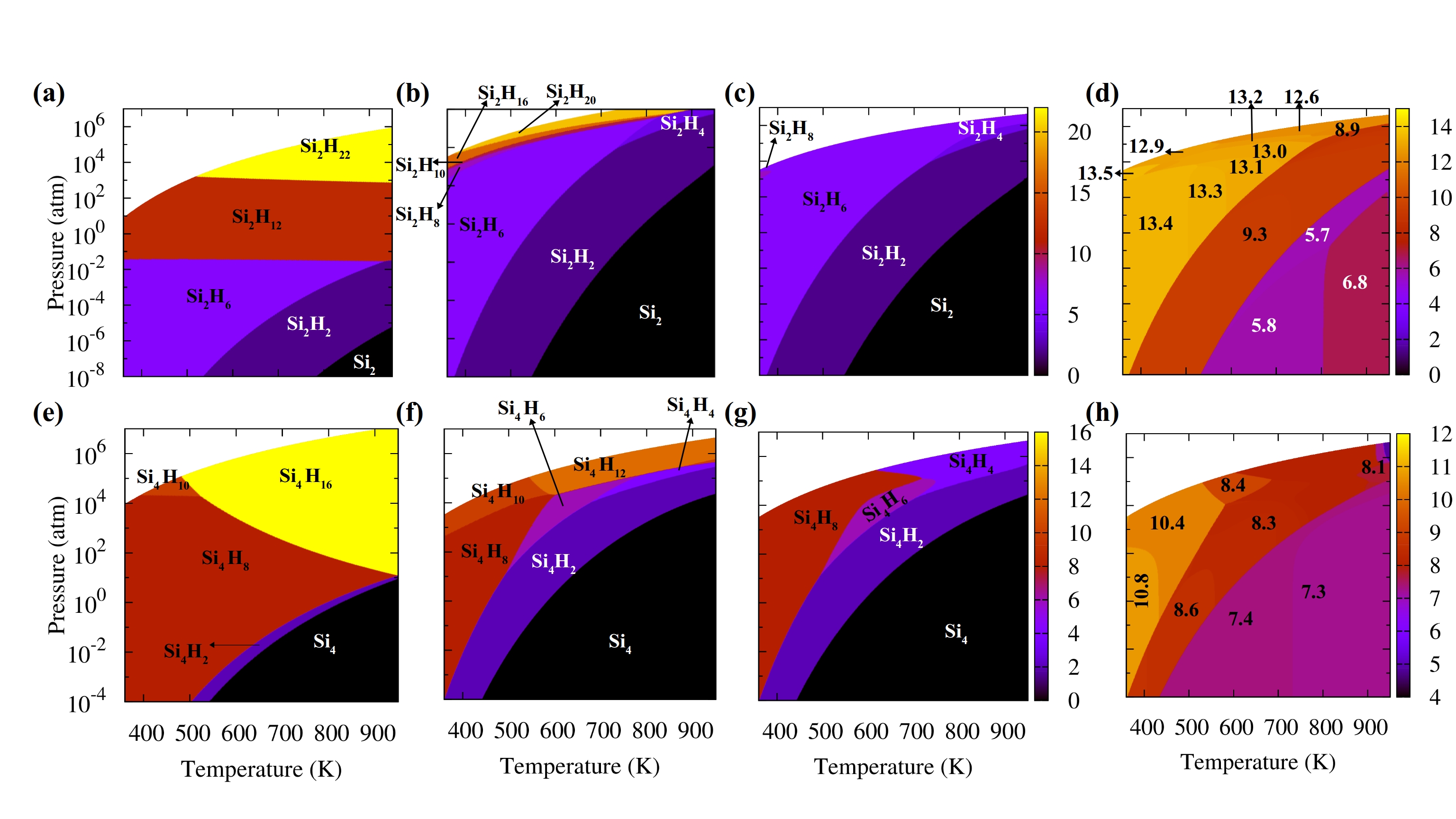}
    \caption{Phase diagrams of Si$_\mathrm{2}$ with H$_\mathrm{2}$ reactive gas phase calculated by (a) aiAT@REGC (b) MBAR@REGC. MBAR@REGC phase diagrams of (c) chemisorbed Si$_\mathrm{2}$H$_N$ and (d) HOMO-LUMO gap of Si$_\mathrm{2}$H$_N$. Phase diagrams of Si$_\mathrm{4}$ with H$_\mathrm{2}$ reactive gas phase calculated by (e) aiAT@REGC and (f) MBAR@REGC. MBAR@REGC phase diagrams of (g) chemisorbed Si$_\mathrm{4}$H$_N$ and (h) HOMO-LUMO gap of Si$_\mathrm{4}$H$_N$ at PBE0 level. HOMO-LUMO gaps in panels (d) and (h) are in eV.}
    \label{fig:pd2}
\end{figure*}
To better characterize the structure of the adsorbate layers, in Figs. \ref{fig:lj-rdf1}a--b and Figs. \ref{fig:lj-rdf2}a--c we show the $g_{xy}(r)$, i.e., the radial distribution functions (RDF) in the $xy$-plane for the different adsorbate layers (i.e., for B-particles in a slab $z_0 \pm \delta z_0$ as specified in each panel). The structures shown in Fig. \ref{fig:lj-rdf2}b--c are obtained via MBAR by evaluating Eq. \ref{eq:avegr}. We observe that the first monolayer and second monolayer (I) $N_\textrm{B} = 45$ have a $g_{xy}(r)$ characteristic of the solid phase with well-defined peaks and long-range order, whereas for the second monolayer (II) $N_\textrm{B} = 59$, the $g_{xy}(r)$ is more disordered. In the relaxed structure of $\textrm{A}_9\textrm{B}_{59}$ (Fig. \ref{fig:lj-rdf2}a), B particles occupy approximately both hollow and bridge sites, relative to the top A$_9$ layer and form a ring-like structure around the projection of the A particles. At (200 K, $2.03 \times 10^{-17}$ atm), the average radial distribution function $\langle g_{xy}(r) \rangle $ of this phase shares some similar peak positions with that of its lowest-energy isomer. It is clear that the ring structure formed by B particles can be still found in the average adsorbate structure though there are a few B particles diffusing around the projection of the A particles. At (600 K, $8.89 \times 10^{-2}$ atm), more and more B particles diffuse and the ring structure is not as noticeable as before. Consistently, the $\langle g_{xy}(r) \rangle$ shares a few major peaks with that of its lowest-energy isomer, but they appear more smeared. \\ The example of $\langle g_{xy}(r)\rangle $ at the two state points was selected in order to demonstrate the power of the REGC sampling to reveal detailed thermodynamic information on the simulated system. A crucial observation is that such information is already contained in the REGC sampling, no further simulation is needed, only post-processing statistical analysis of the sampled data points is required. \\
Coming back to the differences between aiAT@REGC and MBAR@REGC phase diagrams (Fig.~\ref{fig:pd1}), we observe, in Fig.~\ref{fig:lj-sdf} that up to the complete first monolayer ($N_\textrm{B}=18$), the adsorbed particles have essentially no freedom to move in the $z$ direction. As soon as the second monolayer is established, the adsorbed particles display a broader and broader distribution along the $z$ direction. The distribution becomes even bimodal for 
$N_\textrm{B}\ge 62$. This enhanced configurational freedom creates a large, negative, vibrational free energy contribution that stabilizes the higher coverages compared to when only the energetic contribution is taken into account (as in the aiAT@REGC phase diagram).

\subsection{\label{ssec:si-clus} \textit{Ab initio} \ce{Si2H_$N$}  and \ce{Si4H_$N$}clusters}
The REGC algorithm coupled to \textit{ab initio} MD was applied to identify the thermodynamically stable and metastable compositions and structures of Si$_M$H$_N$ ($M$=2, 4) clusters at realistic temperatures and pressure of the molecular hydrogen gas.

\subsubsection{\label{ssec:sipd}Phase diagram}
\paragraph{Si$_2$} Twenty replicas of Si\textsubscript{2} are selected in contact with different thermodynamic states, that is, with temperatures of 500, 650, 800, and 950 K and \ce{H2} chemical potentials of -0.2, -0.16, -0.12, -0.08, and -0.05 eV. The selection of the temperature range is made according to the experimental deposition temperature of chemical vapor deposited silicon films \cite{doi:10.1063/1.347215,1347-4065-30-2R-233}, which starts from around 600 K. Ideally, the lowest $\mu_{\ce{H}}$ should be around $-1.2$~eV, which is the half adsorption energy of H$_2$ on Si$_2$, according to our DFT calculations (see details in appendix). However, in order to focus the sampling on a more interesting region, where more H atoms are adsorbed, we started from a much higher minimum $\mu_{\textrm{H}_2}$. The studied Si$_{2,4}$H$_N$ systems are confined in a sphere with radius 4 {\AA}, by applying reflecting boundaries. This avoids that H atoms diffuse at arbitrary distance from the Si$_M$ cluster, without perturbing the statistics as the cutoff distance is such that the H atoms are not any more interacting with the Si cluster. \textit{Ab initio} molecular dynamics is performed for each system after exchanging particle with the reservoir or swapping with neighboring replicas. For this  REGCMD study, $x_0$ is chosen as 0.9. 
For comparison, we analyzed the stability of Si$_{\mathrm{2}}$H$_N$ clusters using \textit{ab initio} atomistic thermodynamics (aT) in Fig. \ref{fig:pd2}a. For each number of adsorbed hydrogens $N_\textrm{H}$, the lowest DFT energy isomer is identified among all the configurations obtained along the REGC \textit{ab initio} MD sampling. The Gibbs free energy of each phase is calculated as:
\begin{equation}
\Delta G_f(T,p_{\ce{H2}}) = F_{\ce{Si_{2,4}}\ce{H}_N}-F_{\ce{Si_{2,4}}}-N\mu_{\ce{H}}(T,p_{\ce{H2}})
\end{equation}
Here, $F_{\ce{Si_{2,4}}\ce{H}_N}$ and $F_{\ce{Si}_{2,4}}$ are the Helmholtz free energies of the Si$_{\mathrm{2,4}}$H$_N$ and the pristine Si$_{\mathrm{2,4}}$ cluster (at their configurational ground state), respectively. $\mu_{\textrm{H}_2}$ is the chemical potential of the hydrogen molecule. $F_{\ce{Si_2}\ce{H}_N}$  and $F_{\ce{Si_2}}$ are calculated using DFT information and are expressed as the sum of DFT total energy, DFT vibrational free energy in the quasi-harmonic approximation, as well as translational, and rotational free-energy contributions. The dependence of $\mu_{\ce{H_2}}$ on $T$ and $p_{\ce{H_2}}$ is calculated using the ideal (diatomic) gas approximation with the same DFT functional as for the clusters.\cite{beret2014reaction, bhattacharya2013stability, bhattacharya2014efficient} So $p_0$ here is calculated as follows:
\begin{equation}
p_0 = [(\frac{2 \pi m}{h^2})^{\frac{3}{2}}(k_{\ce{B}}T)^{\frac{5}{2}}(\frac{8 \pi^2 I_{\ce{A}} k_{\ce{B}}T}{h^2})\frac{e^{(\frac{k_{\ce{B}}T}{E_{\ce{DFT}}})}}{e^{(\frac{hv_{\ce{HH}}}{k_{\ce{B}}T})-1}}]
\end{equation}
$E_{\ce{DFT}}$ is the DFT total energy, $m$ is the mass, $I_{\ce{A}}$ is the inertia moments, $v_{\ce{HH}}$ is the H-H stretching frequency of 3080 $\textrm{cm}^{-1}$, and \ce{E_{DFT}} of -31.74 eV. The ($p_{\ce{H_2}}, T$) phase diagram of Si$_2$H$_N$ cluster is also constructed via the MBAR@REGC method. As shown in Fig. \ref{fig:pd2}b, besides Si$_2$, Si$_2$H$_2$, and Si$_2$H$_6$, which have their wide stability regions revealed in both phase diagrams, there is a narrow $(T,p_{\ce{H_2}})$ stability domain for Si$_2$H$_4$, which is only revealed by the MBAR@REGC phase diagram that includes without approximation all the anharmonic contributions to the free energy. Another difference between two phase diagrams is that the stable ($p_{\ce{H_2}}, T$) range of each phase is quite different. 
The Si$_2$H$_N$ phases in Fig. \ref{fig:pd2}b include not only chemically adsorbed H atom, but also H$_{2}$ molecule or isolated H atoms. In order to further investigate the chemisorbed phase stability, we construct the phase diagram (Fig. \ref{fig:pd2}c) for a new observable: the number of adsorbed H atoms. A H atom is considered adsorbed on the Si cluster when the distance to the closest Si is smaller than 1.7 \AA.

\paragraph{Si$_4$} Twenty thermodynamic states for the Si$_4$H$_N$ system are selected, with temperature of 560, 685, 810, and 935 K, and chemical potentials of -0.3, -0.2, -0.17, -0.14, and -0.11 eV. The lowest value of $\mu_{\textrm{H}}$ is selected as a bit larger than the half adsorption energy (-0.6 eV) of $\ce{H_2}$ on Si$_4$.
The other settings are the same as in Si$_2$ simulation. 
As for the Si$_2$H$_N$ case, we construct both the aiAT@REGC and MBAR@REGC phase diagram, for comparison, plus the MBAR@REGC phase diagram for the adsorbed H atoms.
In Fig \ref{fig:pd2}e and \ref{fig:pd2}f, the results indicate that two stable Si$_4$H$_4$ and Si$_4$H$_6$ are missing in aT phase diagram. Si$_4$H$_4$ and Si$_4$H$_6$ have considerable larger stable range in chemisorbed phase diagram shown in Fig \ref{fig:pd2}g than in both physi- and chemisorbed one. Besides, the stable ($p_{\ce{H_2}}, T$) range of each phase transitions are quite different in phase diagrams calculated by two method. 
\begin{figure}[h]
   \centering  
    \includegraphics[trim={0.08cm 0.03cm 0cm 0.06cm}, clip, width=0.5\textwidth]{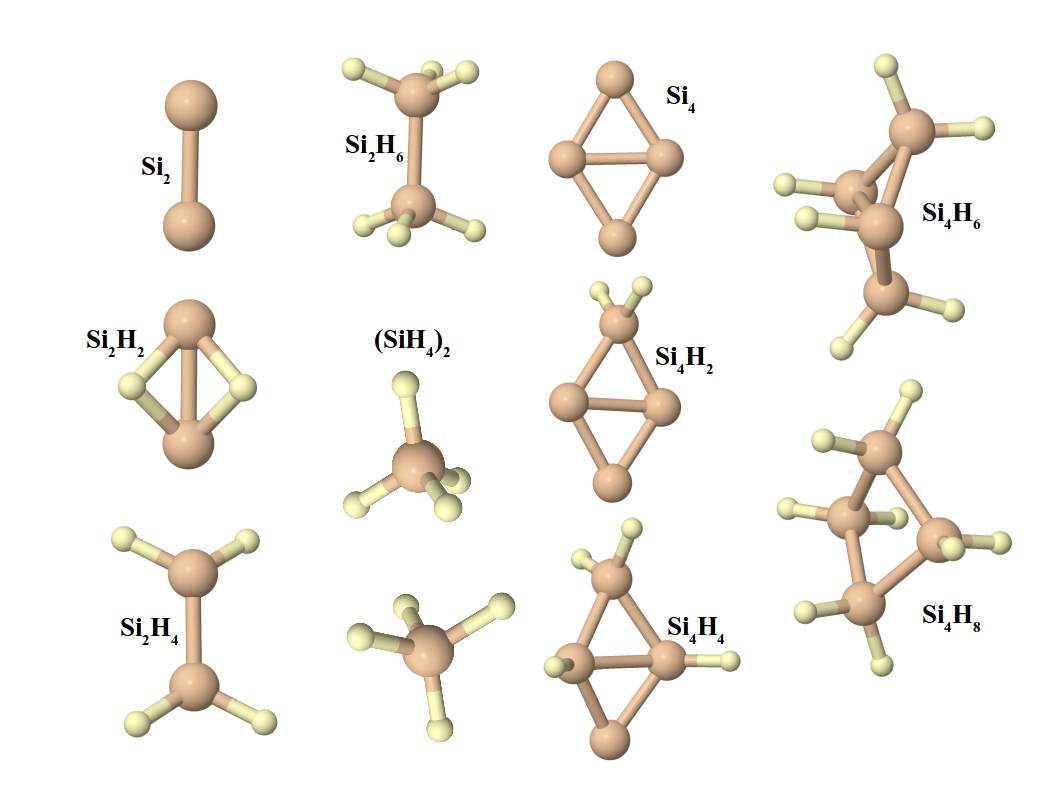}
    \caption{Structures of Si$_\mathrm{2}$H$_N$ and Si$_\mathrm{4}$H$_N$, found by the REGC sampling, that have a region of thermodynamic stability in the phase diagrams of Fig. \ref{fig:pd2}.}
    \label{fig:structs}
\end{figure}

\subsubsection{\label{ssec:sis}Structural and electronic properties of silicon hydrides}
In Fig. \ref{fig:structs}, we show the structures of each thermodynamically stable cluster size appearing in the phase diagrams. All previously reported structures are found in our REGC \textit{ab initio} MD simulations and illustrated in Fig. S2. 
Besides, we identified many other isomers at each composition, via the REGC \textit{ab initio} MD sampling, as shown in Fig. S2.

The HOMO-LUMO gap $E_\textrm{g}$ is also chosen as further observable for the evaluation of phase diagrams for Si$_\mathrm{2}$H$_N$ Fig. \ref{fig:pd2}d and Si$_\mathrm{4}$H$_N$ Fig. \ref{fig:pd2}h. $E_\textrm{g}$ is evaluated as the difference between the vertical electron affinity (VEA) and vertical ionization potential (VIP). The VEA (VIP) is evaluated ---via the PBE0 hybrid \cite{doi:10.1063/1.478522} xc functional, with the Tkatchenko-Scheffler\cite{PhysRevLett.102.073005} pairwise vdW correction--- as the energy difference between the neutral cluster and its monovalent anion (cation), at fixed geometry of the neutral species . It has been clearly shown Fig. \ref{fig:pd2}d and \ref{fig:pd2}h that the HOMO-LUMO gap increases with increasing $N_\textrm{H}$ for both Si$_\mathrm{2}$H$_N$ and Si$_\mathrm{4}$H$_N$, as the VEA decreases with increasing $N_\textrm{H}$ (Fig. S1b and S1e) while VIP increases (Fig. S1c and S1f). This electronic-structure phase diagram can be used to provide guidance to synthesize the material with desired electronic properties, by tuning the environmental conditions, i.e, the temperature and pressure of reactive gas phase.

\section{\label{sec:concl}CONCLUSION AND OUTLOOK}
In summary, we have developed a massively parallel Replica-Exchange Grand-Canonical Monte Carlo/\textit{ab initio} Molecular-Dynamics (REGCMC/MD) algorithm to perform simulations on surfaces/nanoclusters in contact with reactive $(T,p)$ gas and demonstrated how it can be used, in combination with the multistate-Bennet-acceptance-ratio (MBAR) reweighting approach to determine $(T,p)$ phase diagrams. This massively parallel algorithm requires no prior knowledge of the phase diagram and takes only the potential energy function together with the desired $\mu$  and $T$ ranges as inputs. The particle insertion/removal Monte Carlo move, which implements the GC sampling, together with the exchange of configurations among thermodynamic states introduced by RE, allows for an efficient sampling of the configurational space. The approach is appled to an a model surface described by the Lennard-Jones empirical force-fields and small Si clusters in reactive \ce{H2} atmosphere described at the \textit{ab initio}  DFT level. Besides free-energy $(T,p)$ phase diagrams, the combination of the REGC sampling and \textit{a posteriori} analysis via MBAR allows for the determination of phase diagrams for any (atom position dependent) observables, therefore indicating how to tune the environmental condition ($T$ and $p$) to get a material with desired properties. It can therefore be applied to a wide range of practical issues, e.g., dopant profiles, surface segregation, crystal growth and more. Such as an undertaking has its limitation in the cost of \textit{ab initio} molecular dynamics needed for the REGC sampling. However, its embarrassingly parallel nature makes our approach ``towards exascale'' friendly, and can be regarded as a very efficient and internally consistent high-throughput approach. An obvious and indeed currently investigated generalization of the method is to consider more than one reactive gas in the so-called ``constrained equilibrium'' \cite{PhysRevLett.90.046103,PhysRevB.68.045407}(different species do not react in the gas phase, but only at the surface). In order to avoid a dimensional explosion, an algorithm with an adaptive $\mu_i$ grid is under development.

\section{\label{sec:acknow}ACKNOWLEDGEMENTS}
We thank Fawzi R. Mohamed for crucial help in the parallel implementation of the REGC algorithm and helpful discussions. We thank Chunye Zhu for helpful discussions and a critical reading of the manuscript.

\bibliographystyle{apsrev4-1}
%
\section{\label{sec:app}APPENDIX}
\paragraph{Force-field calculations}
The interaction between particles in the surface and gas phase was taken to be a Lennard-Jones 12-6 potentials  $\phi(r)=4\epsilon[(\sigma / r)^{12}-(\sigma / r)^6]$. The parameters $\epsilon_{AB}$, and $\epsilon_{BB}$ are 0.66 and 0.01 eV, respectively. The $\sigma_{AA}$,  $\sigma_{AB}$, and $\sigma_{BB}$ are 2.5, 1.91 and 1.2 \AA. The length of the lattice vectors of this 2D hexagonal supercell is 11.489 {\AA}.
\paragraph{First-principles calculations}
All DFT calculations were performed with the all-electron, full-potential electronic-structure code package FHI-aims\cite{BLUM20092175}. We used the Perdew-Burke-Ernzerhof (PBE) \cite{PhysRevLett.78.1396} exchange-correlation functional, with a tail correction for the van der Waals interactions (vdW), computed using the Tkatchenko-Scheffler scheme \cite{PhysRevLett.102.073005}. A ``tier 1'' basis for both Si and H with ``light'' numerical settings were employed. All AIMD (Born--Oppenheimer) trajectories between REGC attempted moves (0.02 ps each) are performed in the $NVT$ ensemble. The equations of motion were integrated with a time step of 1 fs using the velocity-Verlet algorithm \cite{PhysRev.159.98}. The stochastic velocity rescaling thermostat was adopted, with a decay-time parameter $\tau\,=\,0.02$~ps, to sample the canonical ensemble\cite{doi:10.1063/1.2408420}. The reflecting conditions to confine the system in a sphere of radius 4 \AA~are imposed via PLUMED \cite{BONOMI20091961} interfaced with FHI-aims, by applying a repulsive polynomial potential of order 4.
\end{document}